\documentstyle[12pt]{article}
\newcommand{\be}{\begin{eqnarray}}
\newcommand{\ee}{\end{eqnarray}}

\title{\begin{flushright}
{\normalsize NUC-MINN-98/10-T\\
November 1998 \\}
\end{flushright}
\vspace*{0.3in}
{\bf HYDROGEN-LIKE ATOMS FROM ULTRARELATIVISTIC NUCLEAR COLLISIONS}}
\author{{\bf Joseph Kapusta}$^{\dagger}$ and
 {\bf Agnes Mocsy}$^{\ddagger}$\\
  {\it School of Physics and Astronomy}\\
   {\it University of Minnesota}\\
   {\it Minneapolis, MN 55455}}

\date{}

\begin{document}

\maketitle
\begin{abstract}

The number of hydrogen-like atoms produced when heavy nuclei collide is
estimated for central collisions at the Relativistic Heavy Ion Collider
using the sudden approximation of Baym {\it et al}.
As first suggested by Schwartz, a simultaneous measurement of the
hydrogen and hadron spectra will allow an inference of the
electron or muon spectra at low momentum where a direct experimental
measurement is not feasible.\\

\end{abstract}
%\vspace{2.0in}

\vspace*{0.2in}

\noindent
$^{\dagger}$ kapusta@physics.spa.umn.edu\\
$^{\ddagger}$ amocsy@nucth1.spa.umn.edu

\newpage

The production rate of lepton pairs is a rapidly increasing function
of temperature and so has long been considered a good probe of the
initial high energy density phase of ultrarelativistic nuclear
collisions \cite{qm96}.  The experimental detection of such direct
leptons is a problem in the sub-GeV range of transverse momentum
due to the large number of charged hadrons produced and the
need to disentangle direct leptons from those arising from
hadron decays.  But this is just the kinematic range characterizing
a quark-gluon plasma at a temperature of 200 to 500 MeV.

Schwartz \cite{Mel} proposed to measure
the distribution of atoms formed by the binding of
a directly produced lepton to one of the charged hadrons emerging
from the final state of the nuclear collision.  A measurement of
the charged hadrons and of the atoms, together with a theoretical
calculation relating the distributions of the three particle species,
would then imply the spectrum of leptons.
The beauty of the idea lies in the fact that nearly all indirectly
produced leptons arise from the decay of hadrons, and these decays
occur too long after the collision to allow an atom to be formed.
Of course, one still cannot tell whether the leptons were produced
in quark-gluon plasma or in hadronic matter, but this is another issue.

Five years ago Baym, Friedman, Hughes and Jacak calculated the
relationship among the spectra of the atoms and of the charged hadrons
and leptons which comprise them \cite{Baym}.  The formula reads:
\begin{equation}
\frac{dN_{\rm atom}}{dy d^2p_{\perp, {\rm atom}}} =
8 \pi^2 \zeta(3) \alpha^3 m^2_{\rm red}
\frac{dN_h}{dy d^2p_{\perp, h}}
\frac{dN_l}{dy d^2p_{\perp, l}} \, .
\end{equation}
Here $\zeta(3) = 1.202...$ and $m_{\rm red}$ is the reduced mass
of the hadron and lepton making up the atom.  Since the binding
energy is so small it is an excellent approximation to evaluate
the hadron and lepton rapidities at the same rapidity as that of the
atom, and to equate their transverse velocities as well:
${\bf p}_{\perp, {\rm atom}}/m_{\rm atom} =
{\bf p}_{\perp, h}/m_h =
{\bf p}_{\perp, l}/m_l$.
This formula is based on the sudden approximation which simply
asks for the overlap of the outgoing wave functions of the hadron
and the lepton with their hydrogenic state.  The sudden approximation
is valid because these particles are formed in a nuclear volume
which is extremely small in comparison to the size of the hydrogen atom
and over a time interval which is extremely small in comparison to
the Bohr period.  The specific focus of Baym {\it et al}. was
on $\pi$-$\mu$ atoms.  Here we shall be interested in p-e, p-$\mu$,
$\pi$-e and $\pi$-$\mu$ atoms.
Our essential contribution is to estimate $dN/dy d^2p_{\perp}$
for the leptons, protons and pions in the relevant range of transverse
momentum, and from these to estimate the number of hydrogenic atoms
to be formed in central Au+Au collisions at the Relativistic Heavy
Ion Collider (RHIC).

First we estimate the number of leptons produced in the quark-gluon
plasma phase.  The reaction rate for the process $q + \overline{q}
\rightarrow l^+ + l^-$ is:
\begin{eqnarray}
R_q &=& 12 \int \frac{d^3p_1}{2E_1 (2\pi)^3} \frac{d^3p_2}{2E_2 (2\pi)^3}
\frac{d^3p_+}{2E_+ (2\pi)^3} \frac{d^3p_-}{2E_- (2\pi)^3}
\nonumber \\
&\times& f_{\rm FD}(E_1) f_{\rm FD}(E_2) \left| {\cal M} \right|^2
(2\pi)^4\delta^4(p_1 + p_2 -p_- -p_+) \, ,
\end{eqnarray}
where the 12 arises from three colors and four possible spin states of the
colliding quarks, $f_{\rm FD}$ is the Fermi-Dirac distribution,
and ${\cal M}$ is the matrix element for the reaction. Approximating
$f_{\rm FD}(E) \approx e^{-E/T}$ the momentum distribution for
negatively charged leptons becomes \cite{KKMM}:
\begin{eqnarray} \label{rate}
E_- \frac{d^3R}{d^3p_-} &=& \sum_{q=u,d,s} \frac{e^2 e_q^2}{(2\pi)^6}
\frac{T}{2 E_-} e^{-E_-/T} \int ds
\ln\left(1+e^{-s/4 E_- T}\right) \nonumber \\
&\times& \left[ 1 + 2(m_q^2+ m_l^2)/s +4m_q^2 m_l^2/s^2 \right]
\sqrt{(1 - 4 m_l^2/s)(1-4m_q^2/s)} \, ,
\end{eqnarray}
where $s = (p_1 +p_2)^2$.
The expression for positively charged leptons is the same.
If the masses of the leptons and the quarks can be neglected this
simplifies to
\begin{equation}
E_- \frac{d^3R}{d^3p_-} = \frac{\alpha^2}{3\pi^4} T^2
e^{-E_-/T} \, .
\end{equation}
For the range of transverse momentum of interest in this context
the masses of the muon and strange quark cannot be neglected.

To obtain the total number emitted we integrate over the space-time
volume according to Bjorken's model \cite{Bj}.  For central collisions:
\begin{eqnarray}
\frac{dN}{dy d^2p_T} &=& \sum_{q=u,d,s} \left( \frac{e_q}{e}\right)^2
\frac{3\alpha^2}{8\pi^3}
\left(\tau_0 T_0^3 R_T\right)^2 \int_{-\infty}^\infty \frac{d\eta}{E}
\int_{T_c}^{T_0} \frac{dT}{T^6} e^{-E/T}
\int_{s_{\rm min}}^\infty ds \ln\left(1 + e^{-s/4 E T}\right) \nonumber \\
&\times& \left[ 1 + 2(m_q^2+ m_l^2)/s +4m_q^2 m_l^2/s^2 \right]
\sqrt{(1 - 4 m_l^2/s)(1-4m_q^2/s)} \, .
\end{eqnarray}
Here $R_T$ is the nuclear radius, $y$ is the momentum space rapidity,
$\eta$ is the position space rapidity, $T_0$ is the temperature when
the plasma is first considered to be thermalized, and $T_c$
is the critical or phase transition temperature.  In the Bjorken model
the temperature drops with proper time $\tau$ according to
\begin{equation}
T(\tau) = \left( \frac{\tau_o}{\tau} \right)^{1/3} T_0 \, .
\end{equation}
Finally $E = m_{\perp} \cosh(y-\eta)$ and $m_{\perp} = \sqrt{m_{\rm l}^2
+p_{\perp}^2}$ where the z-axis is the beam axis.

In general the integrals must be done numerically.  However, if the
masses can be neglected then the above simplifies to:
\begin{eqnarray}
\frac{dN}{dy d^2p_T} &=& \frac{2 \alpha^2 R_T^2}{\pi^3}
\frac{(\tau_0 T_0^3)^2}{p_T^4} \left[ (p_T/T_0)^3 K_1(p_T/T_0)
+ 2 (p_T/T_0)^2 K_2(p_T/T_0) \right. \nonumber \\
&&\left. -(p_T/T_c)^3 K_1(p_T/T_c) - 2 (p_T/T_c)^2 K_2(p_T/T_c)
\right] \, .
\end{eqnarray}
If one is interested in hydrogen-like atoms with a transverse momentum
of a few GeV/c then the transverse momentum of the lepton must
have been
\begin{equation}
p_{\perp,l} = \frac{m_l}{m_{\rm atom}}
p_{\perp,{\rm atom}} \, ,
\end{equation}
which is just a few MeV/c for electrons and a few hundred MeV for muons.
For electrons the above formula simplifies even more since their
transverse momentum is always much less than the temperature.
\begin{equation}
\frac{dN}{dy d^2p_T} = \frac{\alpha^2 R_T^2}{2\pi^3}
\left(\tau_0 T_0 \right)^2 \left[ \left( \frac{T_0}{T_c} \right)^4
\ln \left(\frac{1.20 T_c}{p_{\perp}} \right)
- \ln \left(\frac{1.20 T_0}{p_{\perp}} \right) \right]
\end{equation}
This invariant distribution diverges logarithmically at small
transverse momentum.

Our estimates of the lepton distributions also include those coming from the
mixed phase as the system goes through a first-order phase transition. The
contribution from the quark-gluon plasma phase at the phase transition
temperature $T_c$ is given by the distribution (\ref{rate}) multiplied by
the volume fraction occupied by the plasma, $f_{\rm plasma} =
\left(r\tau_c/\tau-1\right)/(r-1)$
plus the corresponding rate in the hadronic phase multiplied
by the hadronic volume fraction $f_{\rm had} =1-f_{\rm plasma}$.
Here r is the ratio of the number of degrees of
freedom in the two phases.  The hadronic rate is obtained from the
annihilation process $\pi^+ + \pi^- \rightarrow l^+ + l^-$.  These
calculations are standard and well-known \cite{KKMM}.

The canonical picture of central collisions of gold nuclei at RHIC
is that the central rapidity region will be almost baryon free \cite{Bj}.
However, central collisions of lead nuclei at the SPS \cite{qm96}
and extrapolations of nucleon collisions such as LEXUS \cite{lexus}
suggest that the baryon rapidity distribution may be roughly flat.
Furthermore, initial temperatures at RHIC may be as high as
500 MeV \cite{pcm} and this will be reflected in the final transverse
mass distribution of the outgoing protons since high temperatures
eventually get converted to transverse flow.  Indeed, inverse slopes
for protons for central lead collisions at the SPS already reach
300 MeV \cite{qm96}.  Therefore a not unreasonable estimate for
protons at RHIC is to assume a flat rapidity distribution times
an exponential falloff in transverse mass.
\begin{equation}
\frac{dN_p}{dy d^2p_{\perp}} = \frac{Z}
{2\pi y_0 (m_p+T_p) T_p} \, \exp\left[(m_p-m_{\perp p})/T_p\right]
\end{equation}
$Z$ is the charge of a single nucleus, $2y_0$ is the rapidity
gap between projectile and target nuclei, $m_{\perp}$ is the
proton's transverse mass, and $T_p$ is the proton effective
temperature (not really a temperature, just an inverse slope).
Similarly, we assume the charged-pion distribution to be:
\begin{equation}
\frac{dN_{\pi}}{dy d^2p_{\perp}} = \frac{dN_{\pi}}{dy} \frac{Z}
{2\pi (m_{\pi}+T_{\pi}) T_{\pi}} \, \exp\left[(m_{\pi}-m_{\perp\pi})/
T_{\pi}\right],
\end{equation}
with the pion rapidity distribution $dN_{\pi}/dy = 1000$.

To get a rough idea of the numbers assume: $T_0 = 3T_c = T_p = T_{\pi}
= 480$ MeV, $Z = 79$, $R_T = 7$ fm for a
central collision at the maximum RHIC energy of 100 GeV per nucleon
per beam, $r = 19/3$ for three quark flavors and a hadron gas with
$g_h = 7.5$ effective degrees of freedom \cite{dof}.

In figures 1 and 2 we plot the transverse momentum distributions of
p-e, $\pi$-e, p-$\mu$ and $\pi$-$\mu$ atoms for a central Au+Au
collision at RHIC.  The trends are easily understood on the basis
of eq. (1).  The muonic atoms dominate the electronic ones because
they have a greater reduced mass, and the pionic ones dominate the
protonic ones because charged pions are much more abundant than protons.

In figures 3 and 4 we plot the number of hydrogen-like atoms
expected per unit of rapidity per day for transverse momentum
bigger than $p_{\perp,{\rm min}}$ for Au+Au collisions with an
impact parameter of 1 fm or less.
A beam luminosity of $2\times 10^{26}$/cm$^2$ sec is assumed.
For the most abundant species, $\pi$-$\mu$ atoms, we estimate
about 1000 per unit of rapidity per day for transverse momenta
larger than 1 GeV/c.

In conclusion, we have reinvestigated the rates for the production
of hydrogen-like atoms at RHIC.  The results are quite promising
for their experimental detection.  It remains to be seen whether
an efficient detector can be designed to observe them.

\section*{Acknowledgements}
We are grateful to J. Sandweiss for suggesting that we investigate this
problem.  This work was supported by the U.S. Department of Energy
under grant DE-FG02-87ER40328.

\newpage

\section*{Figure Captions}

\noindent
Figure 1: The transverse momentum distributions for electronic
atoms for a central Au+Au collision at RHIC.\\

\noindent
Figure 2: The transverse momentum distributions for muonic
atoms for a central Au+Au collision at RHIC.\\

\noindent
Figure 3: The number of electronic atoms produced with a transverse
momentum greater than the indicated value per unit of rapidity
per day at RHIC.  These assume design luminosity and impact parameters
less than 1 fm for Au+Au collisions.\\

\noindent
Figure 4: The number of muonic atoms produced with a transverse
momentum greater than the indicated value per unit of rapidity
per day at RHIC.  These assume design luminosity and impact parameters
less than 1 fm for Au+Au collisions.\\

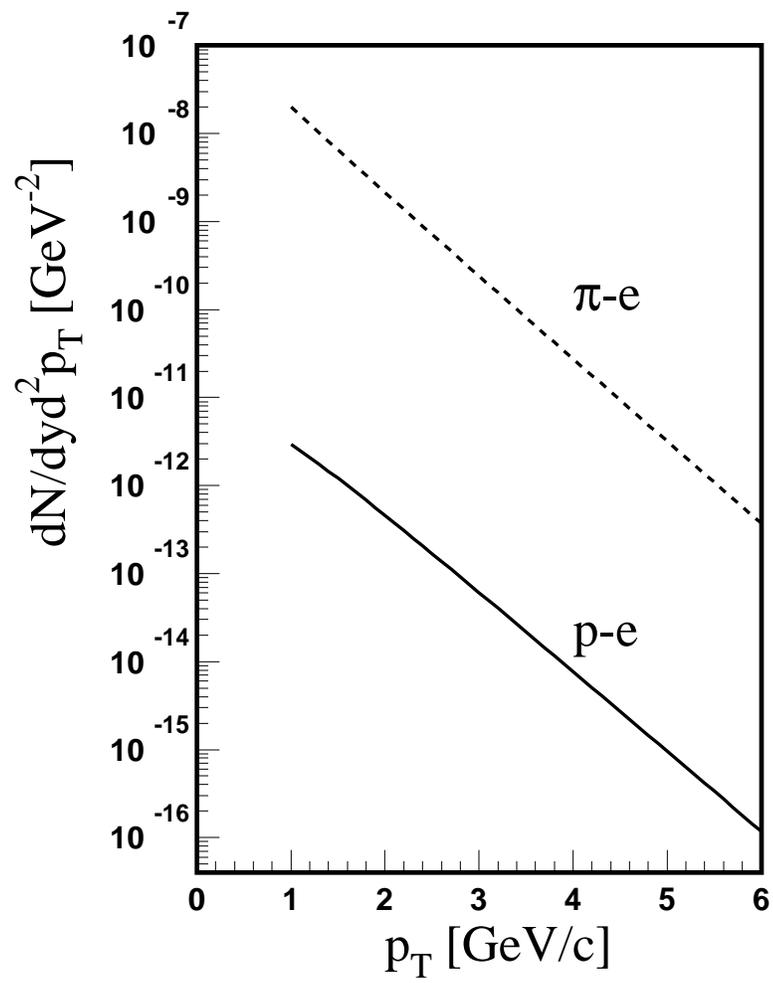
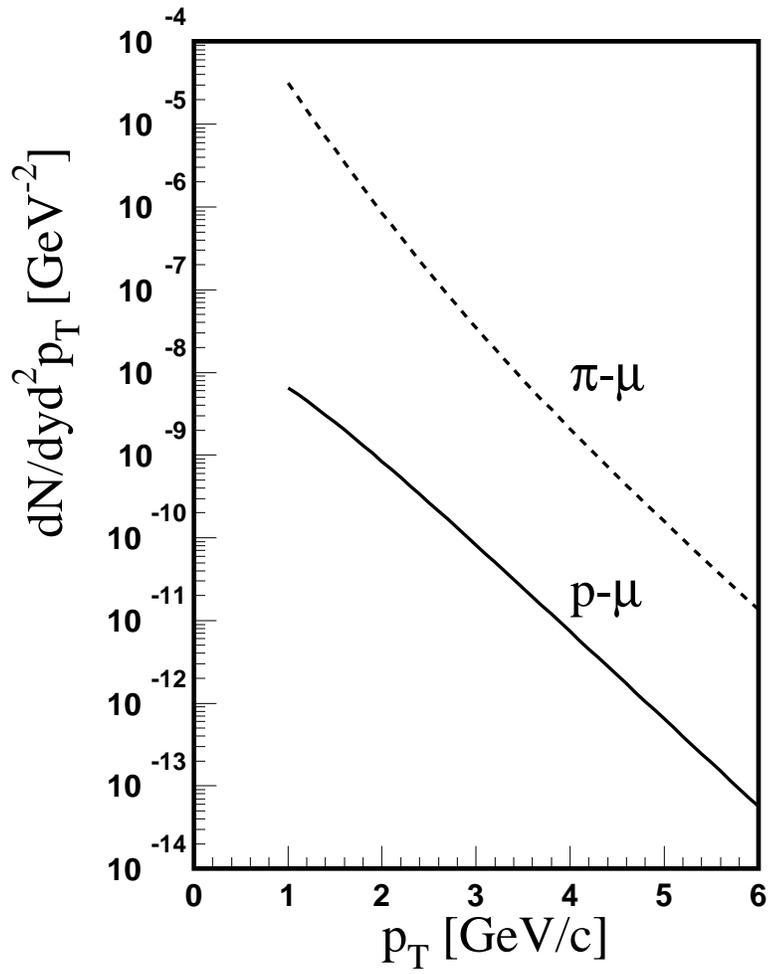
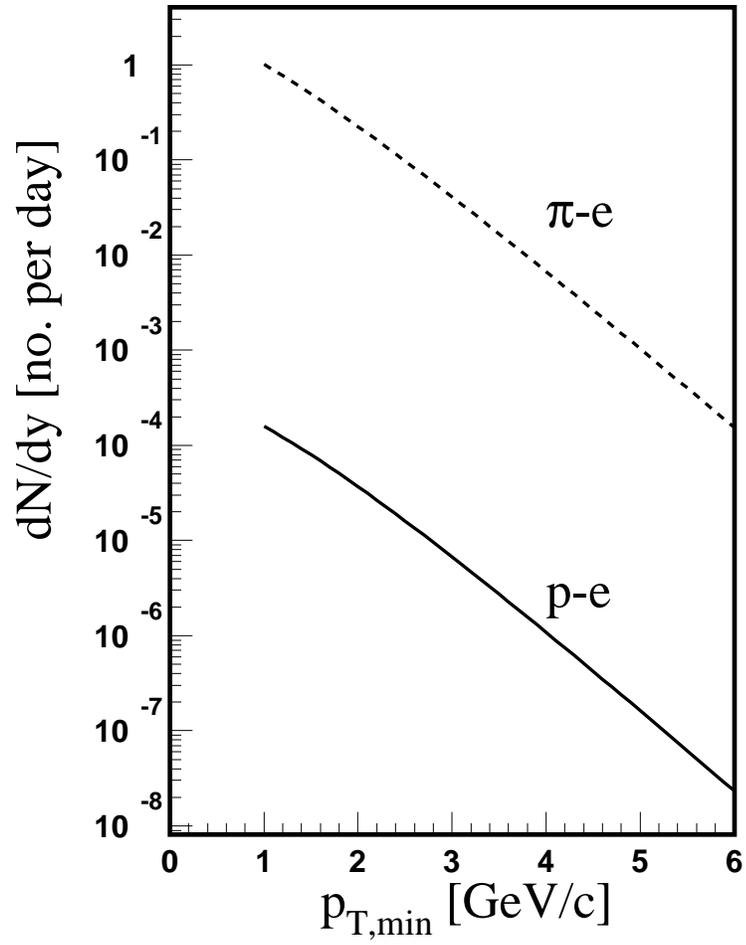
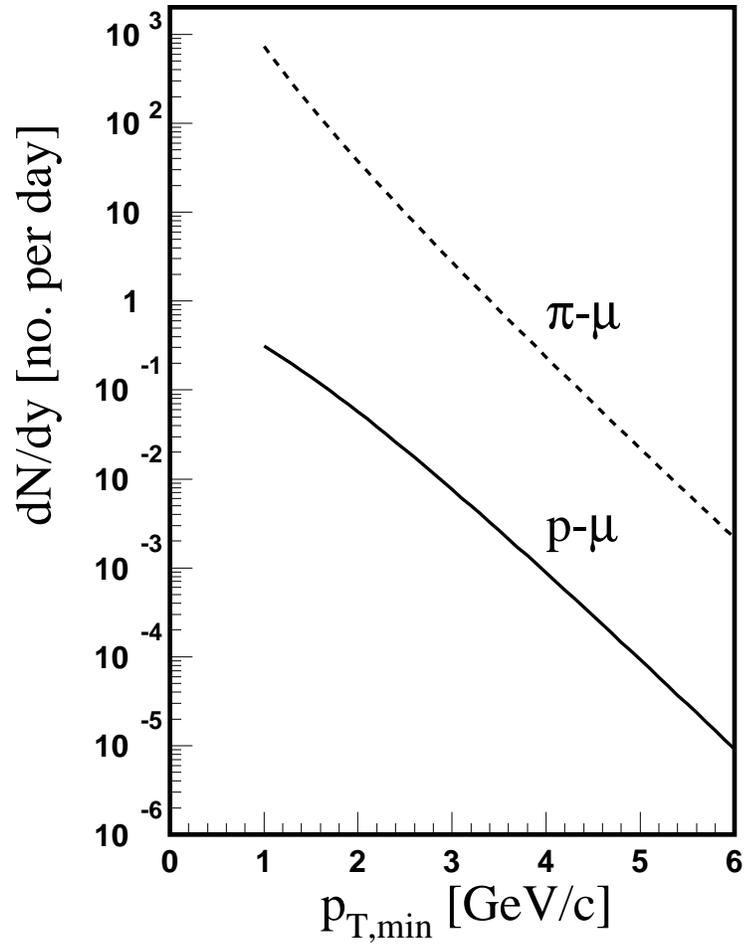
\end{document}